\documentclass[epj]{svjour}
\usepackage{graphics}
\newcommand{\ETslash}{/ \hspace{-.7em} E_T}
\newcommand{\bsmm}{\ensuremath{BR(B_s \to \mu^+\mu^-)}}

\newcommand{\lsim}{\;\raisebox{-.3em}{$\stackrel{\displaystyle <}{\sim}$}\;}

\newcommand{\ssi}{\ensuremath{\sigma^{\rm SI}_p}}

\begin{document}
\title{Supersymmetric  Fits  after the  Higgs Discovery}
\subtitle{and Implications for Model Building}
\author{John Ellis\inst{1} %\and Second author\inst{2}% etc
}                     % Do not remove
\institute{Department of  Physics,  King's~College~London, London WC2R 2LS, United Kingdom
 \and Theory Division, CERN, CH-1211 Geneva 23, Switzerland}
\date{Dec. 19th, 2013}
\abstract{
The data from the first run of the LHC at 7 and 8~TeV, together with the information provided by
other experiments such as precision electroweak measurements, flavour
measurements, the cosmological density of cold dark matter and the direct search
for the scattering of dark matter particles in the LUX experiment,
provide important constraints on supersymmetric models. 
Important information is provided by the ATLAS and CMS
measurements of the mass of the Higgs boson, as well as the negative results of searches at the LHC
for events with $\ETslash$ accompanied by jets, and the LHCb and CMS measurements
of $\bsmm$. Results are presented from frequentist analyses of the parameter spaces
of the CMSSM and NUHM1. The global $\chi^2$
functions for the supersymmetric models vary slowly over most of the
parameter spaces allowed by the Higgs mass and the $\ETslash$ search,
with best-fit values that are comparable to the $\chi^2$ for the Standard Model.
The 95\% CL lower limits on the masses of gluinos and squarks allow
significant prospects for observing them during the LHC runs at higher energies.\\
~\\
KCL-PH-TH/2013-46, LCTS/2013-31, CERN-PH-TH/2013-309 \\
 % end of PACS codes
} %end of abstract
\maketitle
\section{Introduction}
\label{intro}
The discovery of a Higgs boson at the LHC~\cite{H} has given new heart to advocates
of supersymmetry~\cite{Ramond}. Its mass is consistent with the predictions of minimal
supersymmetric models that the lightest Higgs boson should weigh $\lsim 130$~GeV~\cite{susyH}.
Indeed, the measured value of $m_h$ lies in the range where new
physics seems to be required to stabilize the electroweak vacuum~\cite{unstable}, which might
well be supersymmetry~\cite{ER}. Moreover, the measurements of Higgs couplings to other
particles are consistent with the predictions of many supersymmetric models,
which are close to those in the Standard Model. There are no signs so far of
the deviations from the Standard Model couplings that are characteristic of models in
which electroweak symmetry breaking is driven by some new dynamics~\cite{EY}.

On the other hand, neither are there any signs for other types of new
physics, such as might be responsible for dark matter in the form of
massive, weakly-interacting particles whose production could be inferred
in searches for events with jets and missing transverse energy, $\ETslash$ at the LHC.
Supersymmetry with conserved $R$ parity is one such model that
suggests the existence of a dark matter particle that was in thermal
equilibrium in the early Universe and should weigh $\sim 1$~TeV if
it is to have the appropriate cosmological relic density~\cite{Covi}. It is assumed here that the
lightest supersymmetric particle (LSP) that constitutes the dark matter is the
lightest neutralino $\chi$~\cite{EHNOS}, though there are other candidates such
as the gravitino. Important constraints
on such dark matter models are imposed by direct and indirect searches for
dark matter, as well as by LHC searches for $\ETslash$ events, none of
which have found convincing signals~\cite{ETslash}.

Even if $R$ conservation is assumed, the interpretation of all these constraints
is quite model-dependent. For simplicity, we consider here only the minimal
supersymmetric extension of the Standard Model (the MSSM), though there are
well-motivated extensions, e.g., to include any extra singlet superfield (the NMSSM~\cite{Djouadi}).
The MSSM already has over 100 parameters, and it is natural to consider
simplifying hypotheses such as minimal flavour violation (MFV), in which all
flavour violation is related to Cabibbo-Kobayashi-Maskawa mixing~\cite{MFV}. In principle, this model
has 6 additional CP-violating phases~\cite{MCPMFV}, but upper limits on electric dipole moments
offer no suggestion that they are large. Many studies of experimental constraints
focus on versions of the MSSM with MFV in which the soft supersymmetry-breaking
contributions to sfermion, Higgs and gaugino masses, $m_0$ and $m_{1/2}$, 
respectively, as well as trilinear couplings $A_0$, are constrained to be universal 
at some high input scale (the CMSSM)~\cite{CMSSM}, or in generalizations in which the soft
supersymmetry-breaking contributions to Higgs masses are allowed to be
non-universal but equal (the NUHM1)~\cite{NUHM1}. One example of a more restrictive model
is minimal supergravity (mSUGRA), in which the gravitino mass is forced to be
equal to the input scalar mass: $m_{3/2} = m_0$, and the trilinear and bilinear
soft supersymmetry-breaking parameters are related: $A_0 = B_0 + m_0$.

As we shall see, the LHC $\ETslash$ searches impose strong constraints on
models with universal soft supersymmetry-breaking parameters such as the
CMSSM, NUHM1 and mSUGRA, stimulating interest in `natural' models in which the
third-generation squarks are much lighter than those of the first and second
generations, for which experiments give weaker constraints. Also, searches 
for specific $\ETslash$ + jets signatures have been interpreted within simplified 
models in which these topologies are assumed to be the dominant
supersymmetric signatures. There has also been interest in using searches
for $\ETslash$ + monojet, monophoton and mono-$W/Z$ topologies to
look for the direct pair-production of dark matter particles without passing
via the cascade decays of heavier sparticles.

In view of its importance for constraining supersymmetric models, in
Section 2 of this review there is a discussion of Higgs mass calculations
and their uncertainties, as well as indications of their implications for the 
parameter spaces of supersymmetric models. Section~3 presents some
results of global fits~\cite{mc9} to the CMSSM and NUHM1 using the full $\ETslash$ data from
Run~1 of the LHC at 7 and 8~TeV~\cite{ATLAS20}, the measurement by CMS and LHCb of
\bsmm~\cite{bsmm}, and the latest constraints on dark matter
scattering from the LUX experiment~\cite{LUX}. These results include 95\% CL lower limits
on sparticle masses and the prospects for discovering them in Run~2
of the LHC at 13/14~TeV. Section~4 summarizes some pertinent results
within other frameworks such as mSUGRA, `natural' and simplified models.
Finally, Section~5 draws some conclusions for supersymmetric model-building.

\section{The Higgs Mass and Supersymmetry}

As is well known,
the two complex Higgs doublets of the MSSM have eight degrees of freedom,
of which three give masses to the $W^\pm$ bosons and to the $Z^0$ via the
electroweak symmetry breaking, leaving five physical Higgs bosons in the 
physical spectrum: two neutral Higgs bosons $h, H$ that are CP-even (scalar),
one neutral boson $A$ that is CP-odd (pseudoscalar), and two charged bosons $H^{\pm}$.
The tree-level masses of the scalar supersymmetric Higgs bosons are:
\begin{equation}
m^2_{h, H} = \frac{1}{2}\left(m_{A}^2+m_Z^2 \mp 
\sqrt{(m_{A}^2+m_Z^2)^2 -4m_{A}^2m_Z^2\cos^2 2\beta}\,\right)
\label{hH}
\end{equation}
where $\tan \beta$ is the ratio of Higgs v.e.v.s,
from which we see that $m_h$ is bounded from above by $m_Z$~\footnote{This 
upper limit appears because the quartic Higgs
coupling $\lambda$ is fixed in the MSSM to be equal to the square of the electroweak gauge
coupling, up to numerical factors.}.
However, there are important radiative corrections to $m_h$ (\ref{hH})~\cite{susyH},
of which the most important is the one-loop correction due to the top quark and stop squark:
\begin{equation}
\Delta m_h^2=\frac{3m_t^4}{4\pi^2 v^2}\ln\left(\frac{m_{\tilde{t}_1}m_{\tilde{t}_2}}{m_t^2}\right) 
+ \dots \, ,  
\label{deltamh}
\end{equation}
where $m_{\tilde{t}_{1,2}}$ are the physical masses of the stops. We see in (\ref{deltamh})
that the correction $\Delta m_h^2$ depends quartically on the mass of the top, and it implies
that the mass of the lightest Higgs boson may  be as large as
\begin{equation}
m_h \lsim 130~{\rm GeV}.
\label{hmass}
\end{equation}
for stop masses of about a TeV, consistent with the ATLAS and CMS measurements~\cite{H}.

\begin{figure}
\resizebox{0.45\textwidth}{!}{%
  \includegraphics{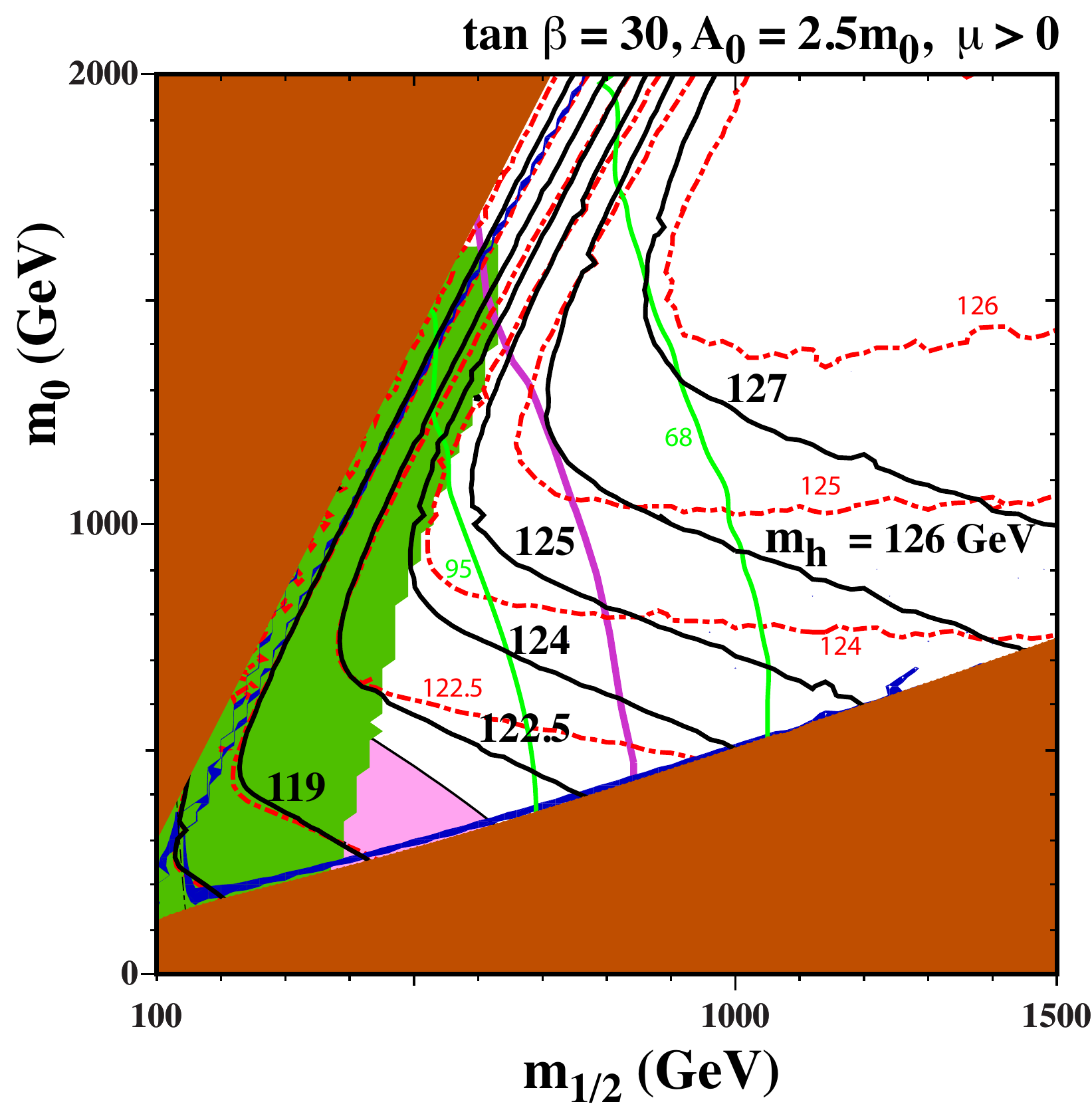}
}
%\vspace{5cm}       % Give the correct figure height in cm
\caption{The allowed regions in the $(m_{1/2}, m_0)$ plane for
$\tan \beta = 30$ and $A_0 = 2.5 m_0$~\cite{EHOW+}. The line styles and shadings are
described in the text. The section of the dark blue coannihilation strip
in the range $m_{1/2} \in (840, 1050)$~GeV
is compatible with the constraints from BR($B_s \to \mu^+ \mu^-$) (green lines marking the
68 and 95\% CL)~\cite{bsmm} and the
ATLAS 20/fb MET search (purple line)~\cite{ATLAS20}, as well as with the LHC $m_H$ measurement.
Good consistency with all the constraints is found if the improved {\tt FeynHiggs~2.10.0} code~\cite{newFH} is used
(black lines): results from a previous version of {\tt FeynHiggs} are indicated by red dotted lines).
}
\label{fig:CMSSM30}       % Give a unique label
\end{figure}

If one wishes to use (\ref{deltamh}) to estimate the stop mass scale, it is clear that the
answer is exponentially sensitive to the Higgs mass, and it is therefore important to
refine the one-loop calculation. Several codes are available that provide complete
two-loop calculations and include the leading dependences of three- and higher-loop
contributions on the strong coupling $\alpha_s$ and the top Yukawa coupling $\alpha_t$. It is also important
to estimate the theoretical uncertainty in the calculation of $m_h$ for given values of the
supersymmetric model parameters, which is typically $\sim 1.5$ to 3~GeV.
In the following results from the {\tt FeynHiggs~2.10.0} code for calculating $m_h$ are used,
which is a significant improvement over previous versions.
As an example of the importance for inferences about the supersymmetric mass scale 
from the measured value of $m_h$, Fig.~\ref{fig:CMSSM30} displays the $(m_{1/2}, m_0)$
plane in the CMSSM for $\tan \beta = 30$, $\mu > 0$ and $A_0 = 2.5 m_0$~\cite{EHOW+}.

The brown shaded wedge at large $m_{1/2}$ and small $m_0$ is excluded
because there the LSP would be the charged ${\tilde \tau_1}$, whereas the lighter
stop, ${\tilde t_1}$, would be the LSP. Adjacent to these wedges are narrow blue strips where
the relic LSP density falls within the range favoured by astrophysics and cosmology.
Measurements of $b \to s \gamma$ exclude
the region shaded green, whereas in the pink region the discrepancy between the
Standard Model and experimental values of the anomalous magnetic moment of
the muon, $g_\mu - 2$, could be explained by supersymmetry~\cite{g-2}. The 95\% CL limit
on $\ETslash$ + jets events at the LHC~\cite{ATLAS20} is represented by the purple line, and the
green lines represent 68 and 95\% CL limits from the value of \bsmm\ measured
by the CMS and LHCb experiments~\cite{bsmm}. Finally, the black lines are contours of $m_h$
calculated with the current version {\tt 2.10.0} of the {\tt FeynHiggs} code~\cite{newFH}, which includes
the leading and next-to-leading $\log(m_{\tilde t}/m_t)$ terms in all orders
of perturbation theory, as calculated using the two-loop
Renormalization-Group Equations (RGEs). The red dashed lines are
calculated with an earlier version of {\tt FeynHiggs} that did not include these
refinements, and we see that the $m_h$ contours diverge significantly at large $m_{1/2}$, in
particular. We also see that there is a region with $(m_{1/2}, m_0) \sim (1200, 600)$~GeV
that is compatible with dark matter and laboratory constraints (except for $g_\mu - 2$)
and corresponds to $m_h \sim 125$~GeV according to the latest version of
{\tt FeynHiggs}, whereas the earlier version would have yielded $m_h < 124$~GeV~\cite{EHOW+}.

Smaller values of $\tan \beta$ would yield smaller values of $m_h$, and larger
values of $\tan \beta$ would be more tightly constrained by \bsmm, though values of
$\tan \beta \lsim 50$ may be compatible with all the constraints. Smaller values
of $A_0$ would also yield smaller values of $m_h$ along the strip near the boundary of the
${\tilde \tau_1}$ LSP wedge where the appropriate dark matter density is obtained, and
this dark matter strip would only extend to lower $m_{1/2}$ in this case. There is a second dark matter
strip close to the boundary with the ${\tilde t_1}$ LSP region, but $m_h$ is too small except
possibly at very large values of $m_0$~\cite{EHOW+}. In general, 
CMSSM models with an LHC-compatible value of $m_h$
do not make a significant contribution to resolving the $g_\mu - 2$ discrepancy~\cite{g-2}.

\section{Global Fits in the CMSSM and NUHM1}

After this first taste of the interplay between the LHC $\ETslash$, $m_h$, \bsmm, dark matter
and other constraints, and their potential implications for models,
I now present some results from a global fit to the relevant data
within the CMSSM~\cite{mc9}. These are compared with the results of a fit within the NUHM1, which offers,
in principle, new ways to reconcile some of the constraints discussed in the previous Section.

These fits are based on a frequentist approach developed by the MasterCode
collaboration~\cite{previous,mc8,others}, and the {\tt MultiNest} tool is used to sample the
CMSSM and NUHM1 parameter spaces~\cite{MultiNest}. The global $\chi^2$ function is calculated including precision
electroweak observables such as $M_W$ and measurements at the $Z^0$ peak, as well as
$g_\mu - 2$. Also included is a full suite of flavour observables such as $b \to s \gamma$
and $B \to \tau \nu$ as well as \bsmm~\cite{mc9}. In addition to the dark matter density, a contribution
from the LUX direct search~\cite{LUX} for the scattering of astrophysical dark matter is also included.

\begin{figure*}
\resizebox{8.5cm}{!}{\includegraphics{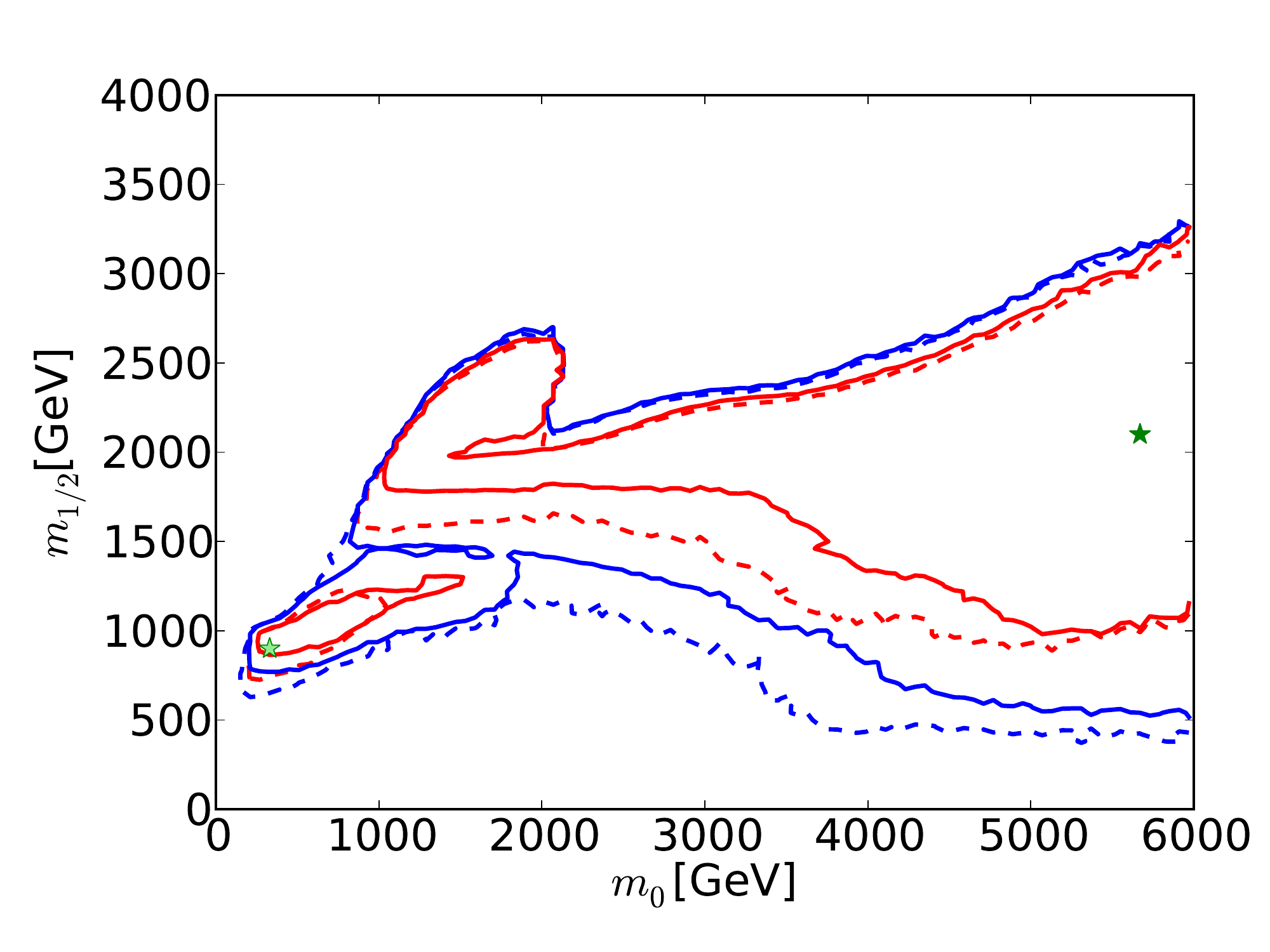}}
\resizebox{8.5cm}{!}{\includegraphics{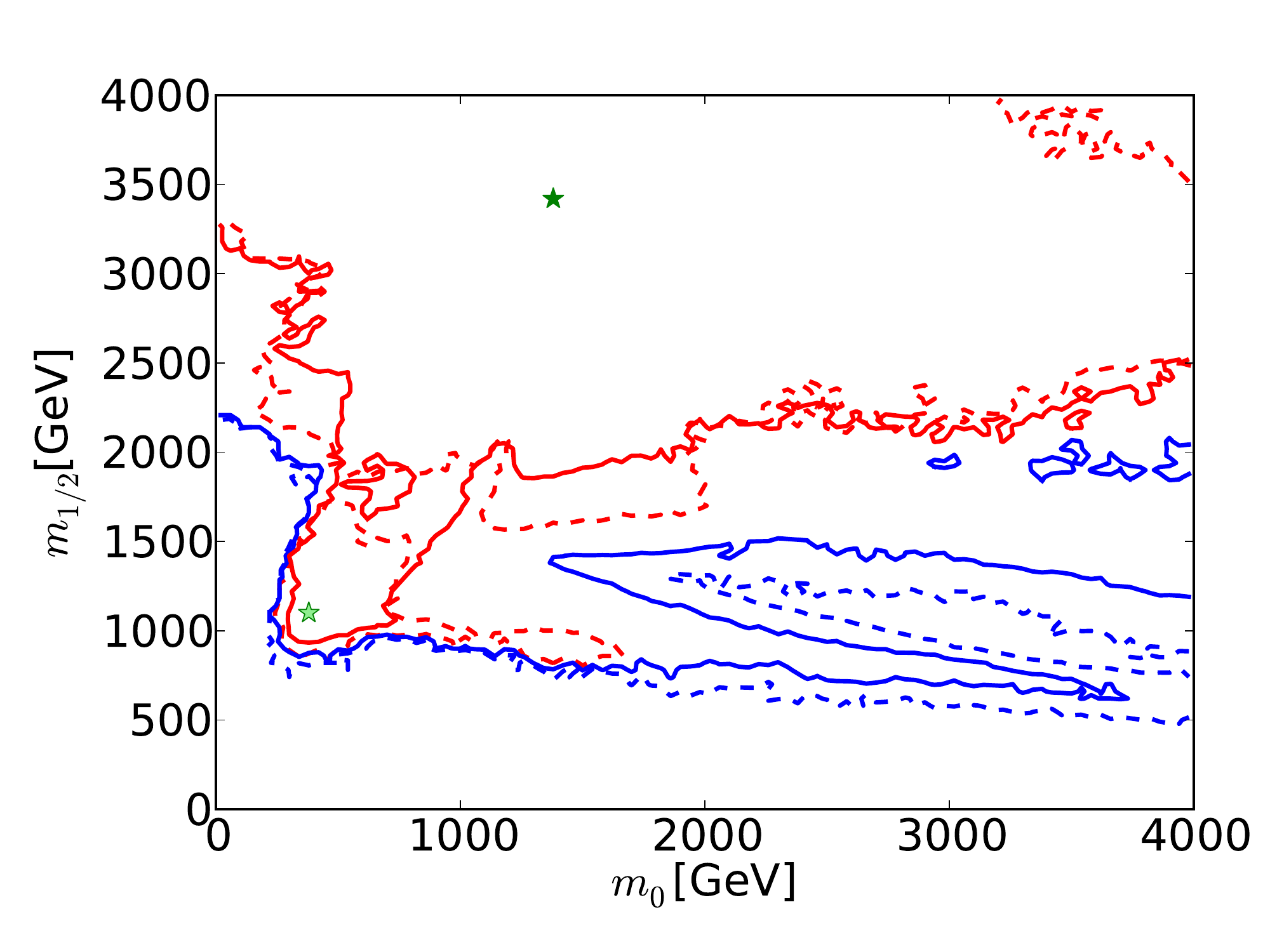}}
%\vspace*{5cm}       % Give the correct figure height in cm
\caption{The $(m_0, m_{1/2})$ planes in the CMSSM
(left), and the NUHM1 (right), after implementing the ATLAS $\ETslash$, \bsmm, $m_h$,
dark matter density, LUX and other relevant constraints~\cite{mc9}.
The results of the current fits are indicated by solid lines
and filled stars, and fits to previous data~\cite{mc8} using the same implementations of
the constraints are indicated by dashed lines and open stars.
The red lines denote $\Delta \chi^2 = 2.30$ contours (corresponding approximately to
the  68\%~CL), and the red lines denote $\Delta \chi^2 = 5.99$ (95\%~CL) contours.}
\label{fig:planes}       % Give a unique label
\end{figure*}

Fig.~\ref{fig:planes} displays $(m_0, m_{1/2})$ planes in the CMSSM (left panel)
and the NUHM1 (right panel), both with $\mu > 0$~\footnote{Results for the CMSSM
with $\mu < 0$ can be found in~\cite{mc9}.}. The best-fit points are indicated by
green stars, the $\Delta \chi^2 = 2.30$ contours that correspond
approximately to the 68\% CL are shown as red lines, and the $\Delta \chi^2 = 5.99$
contours that correspond approximately to the 95\% CL are shown as blue lines. The results 
of the current fit~\cite{mc9} are indicated by solid lines and solid stars, whilst the dashed lines and open stars
represent the results of fits to the data used in~\cite{mc8}, reanalyzed using the current version of {\tt MasterCode}.

In both the CMSSM and the NUHM1, we see two distinct regions: a smaller region around
$(m_0, m_{1/2}) \sim (500, 1000)$~GeV and a larger region extending to larger values
of $(m_0, m_{1/2}$. The low-mass regions correspond to the ${\tilde \tau_1}$
coannihilation strip mentioned in the previous Section, and in the high-mass regions
other mechanisms bring the relic LSP density into the range allowed by astrophysics
and cosmology, notably rapid LSP annihilation via direct-channel $H/A$ resonances
when $m_\chi \sim m_{H/A}/2$, and neutralino-chargino coannihilation, which becomes 
more important when the LSP has a significant Higgsino component. The extra parameter
in the NUHM1 Higgs sectors offers more possibilities for these effects, enabling the
relic density constraint to satisfied at larger values of $m_{1/2}$ and smaller
values of $\tan \beta$ than in the CMSSM~\cite{EHOW+}.

As we see in Table~\ref{tab:bestfits},
he minimum values of $\chi^2$ in the low- and high-mass regions differ by less than
unity in both the CMSSM and the NUHM1. In the case of the CMSSM, the contribution
from $g_\mu - 2$ is smaller in the low-mass region, but the contribution from the ATLAS
jets + $\ETslash$ search is larger. This is also the case in the NUHM1, but other 
observables such as $A_{fb}(b)$ and $A_\ell({\rm SLD})$ also contribute differences
in $\chi^2$ between the low- and high-mass regions that are ${\cal O}(1)$~\cite{mc9}. In general,
the global $\chi^2$ function varies little over much of the $(m_0, m_{1/2})$ planes
explored. Also, the value of $\chi^2$ at the global minimum in the CMSSM is not significantly different
from that in the Standard Model, whereas that in the NUHM1 is $\sim 2$ lower~\cite{mc9}.
The CMSSM and NUHM1 confer no convincing advantages over the Standard Model
in the global fits reported here.

%%%%%%%%%%%%%%%%%%%%%% T A B L E %%%%%%%%%%%%%%%%%%%%%%%%%%%%%%%%%%%%%%%%%
\begin{table*}[!tbh!]
\renewcommand{\arraystretch}{1.5}
\begin{center}
\begin{tabular}{|c|c||c|c|c|c|} \hline
Model & Region & Minimum & $m_0$ & $m_{1/2}$ & $\tan \beta$ \\
  &    & $\chi^2$ & (GeV) & (GeV) & \\ 
\hline \hline
CMSSM  & Low-mass
    & 35.8 & $670$ & $1040$ 
    & $21$ \\
& High-mass & 35.1 & $5650$ & $2100$ & $51$ \\
\hline
NUHM1  & Low-mass
    & 33.3 & $470$ & $1270$ 
    & $11$ \\
& High-mass & 32.7 & $1380$ & $3420$ & $39$ \\
\hline \hline
\end{tabular}
\caption{The best-fit points found in global CMSSM and NUHM1 fits with $\mu > 0$,
  using the ATLAS $\ETslash$ constraint~\cite{ATLAS20}, and the combination of the
  CMS and LHCb constraints on  \bsmm~\cite{bsmm}.
  We list the parameters of the best-fit
  points in both the low- and high-mass regions in Fig.~\protect\ref{fig:planes}.
  The
  overall likelihood function is quite flat in both the CMSSM and the NUHM1, so that the precise
  locations of the best-fit points are not very significant, and we do not quote uncertainties.
  This Table is adapted from~\cite{mc9}.
  }
\label{tab:bestfits}
\end{center}
\end{table*}
%%%%%%%%%%%%%%%%%%%%%% T A B L E %%%%%%%%%%%%%%%%%%%%%%%%%%%%%%%%%%%%%%%%%

Comparing the current fits (solid lines and filled stars) with the results of fits to the
data available in mid-2012 (dashed lines and open stars) reanalyzed with the
current versions of {\tt FeynHiggs} and other codes, we see that the overall
extensions and shapes of the regions allowed at the 95\% CL and favoured at the
68\% CL are quite similar~\cite{mc9}. There is some erosion of the preferred regions at low $m_{1/2}$,
due to the stronger ATLAS jets + $\ETslash$ limit, but the most noticeable features are the shifts to larger masses
of the best-fit points. However, as noted above, the differences between the values
of the global $\chi^2$ function in the low- and high-mass regions are not significant.
The lower-mass regions would require less fine-tuning and hence seem more natural~\cite{ft}.
However, the interpretation of the degree of naturalness is uncertain in the absence of a
more complete theoretical framework.

\begin{figure*}[htb!]
%%%%%%%%%%%%%%%%%%%%%%%%%%%%%%%
\resizebox{8.5cm}{!}{\includegraphics{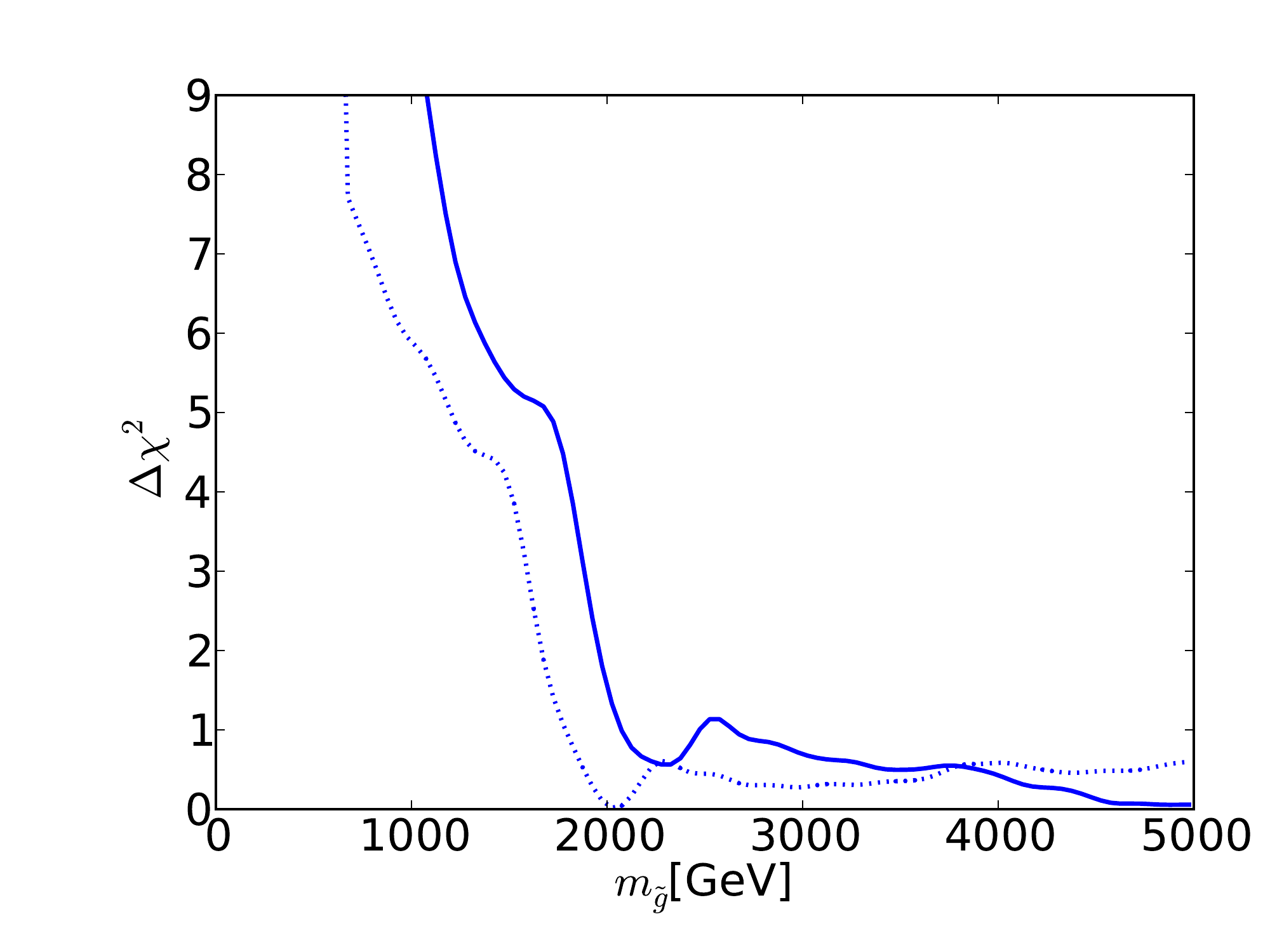}}
\resizebox{8.5cm}{!}{\includegraphics{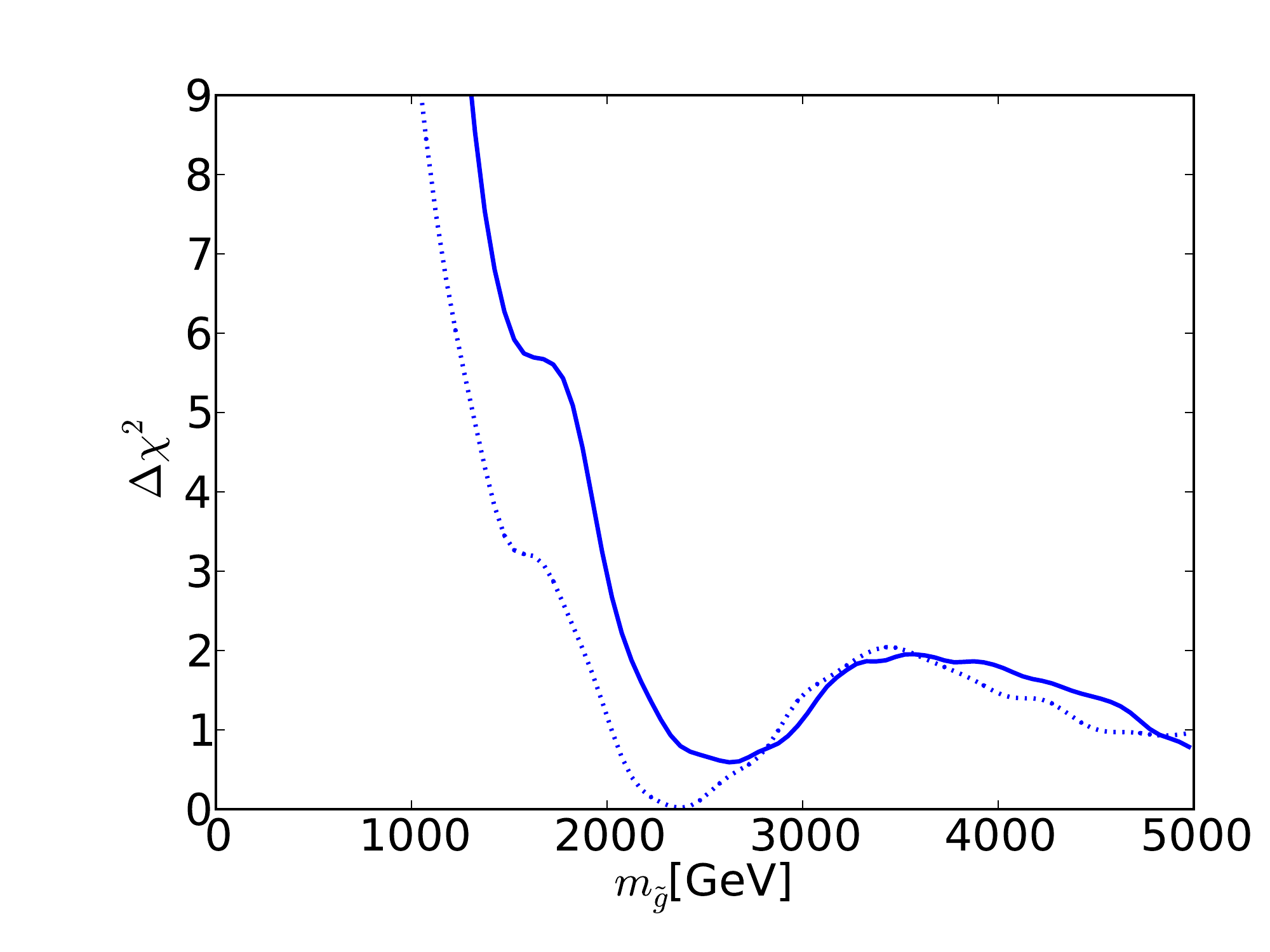}} \\
%%%%%%%%%%%%%%%%%%%%%%%%%%%%%%%
\resizebox{8.5cm}{!}{\includegraphics{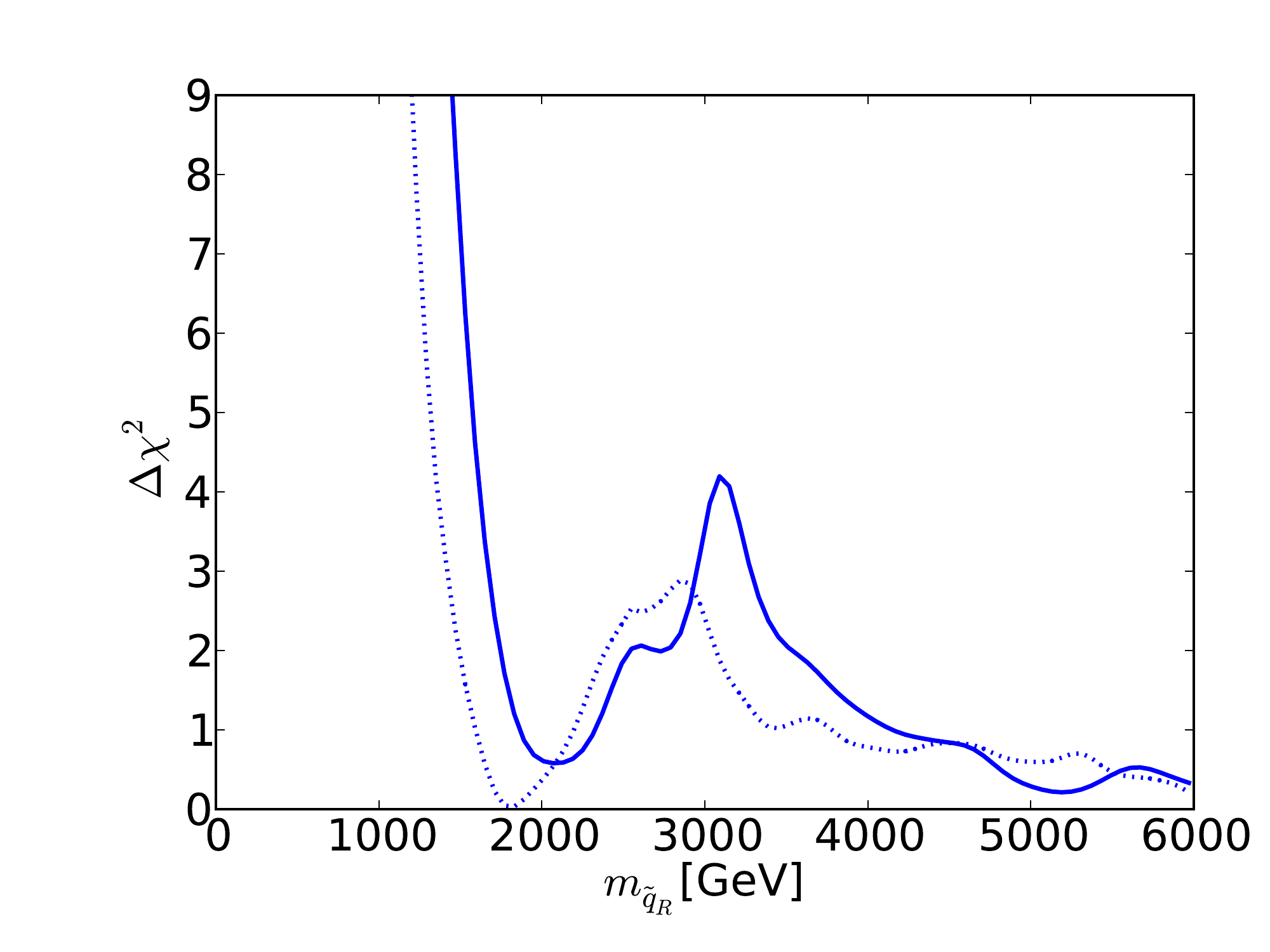}}
\resizebox{8.5cm}{!}{\includegraphics{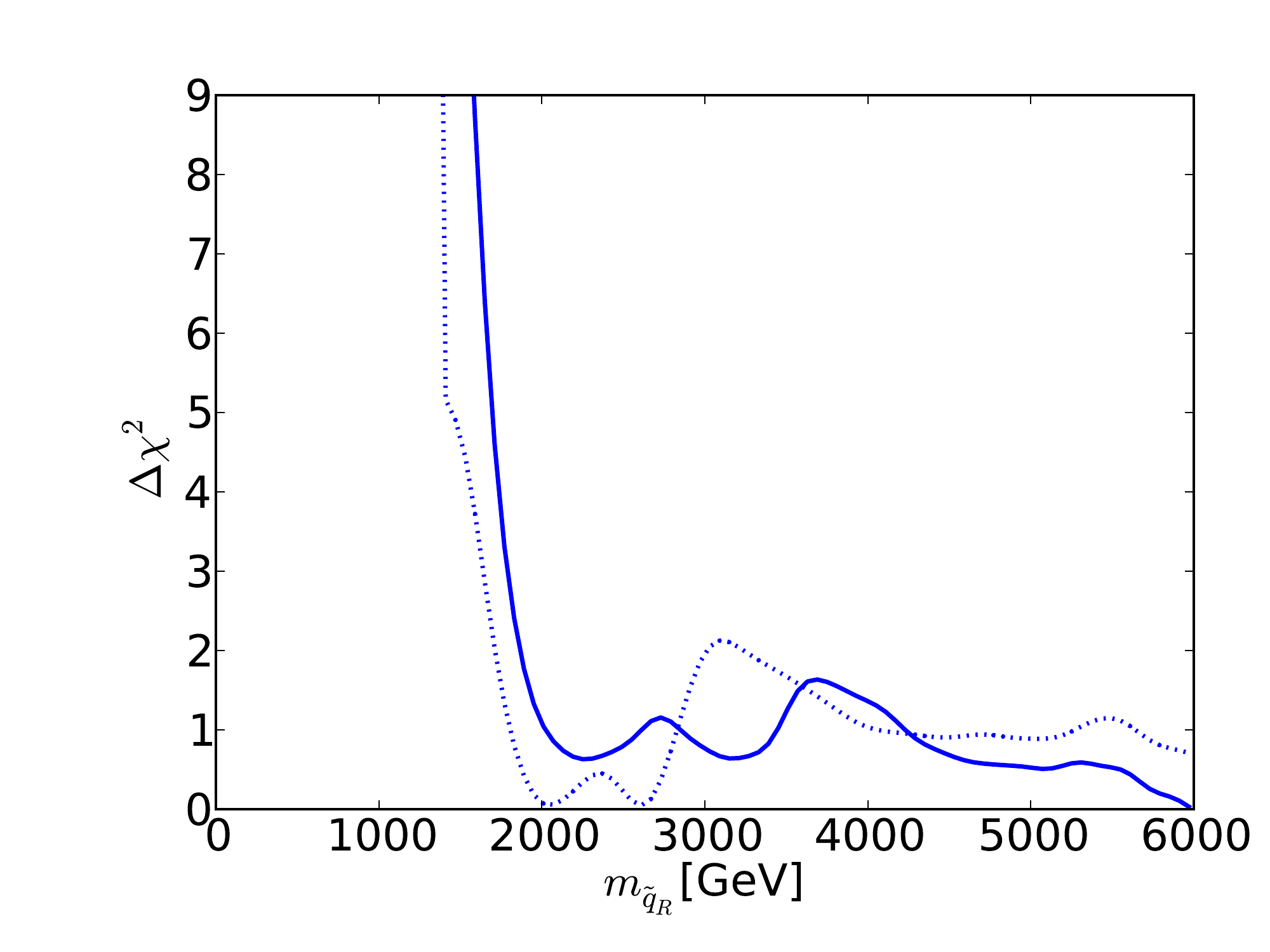}}
%%%%%%%%%%%%%%%%%%%%%%%%%%%%%%%
%\vspace{-1cm}
\caption{The one-dimensional $\chi^2$ likelihood functions in the CMSSM (left) and the NUHM1 (right) for
the gluino mass $m_{\tilde g}$ (upper) and a generic right-handed squark mass $m_{\tilde q_R}$ (lower)~\protect\cite{mc9}.
In each panel, the solid line is derived from a global analysis of the
present data, and the dotted line is derived from an analysis if the data set used
in~\protect\cite{mc8}, using the same implementations of the $m_h$ and dark matter scattering constraints.}
\label{fig:1D}
\end{figure*}

Fig.~\ref{fig:1D} displays the one-dimensional $\chi^2$ functions
for some sparticle masses in the CMSSM (left) and the
NUHM1 (right)~\cite{mc9}. The upper panels are for the gluino mass $m_{\tilde g}$, and the lower
panels are for a generic right-handed squark mass $m_{\tilde q_R}$. The $\chi^2$ function for $m_{\tilde g}$
in the CMSSM falls almost monotonically, whereas the other $\chi^2$ functions exhibit more
structure, corresponding to the structures visible in the $(m_0, m_{1/2})$ planes in
Fig.~\ref{fig:planes}. In each case, the $\chi^2$ functions have been pushed up at
low mass by the ATLAS jets + $\ETslash$ limit, as seen by comparing the solid and dotted lines.

%%%%%%%%%%%%%%%%%%%%%% F I G U R E %%%%%%%%%%%%%%%%%%%%%%%%%%%%%%%%%%%
%%%%%%%%%%%%%%%%%%%%%% F I G U R E %%%%%%%%%%%%%%%%%%%%%%%%%%%%%%%%%%%

\begin{figure*}[htb!]
%%%%%%%%%%%%%%%%%%%%%%%%%%%%%%%
\resizebox{8.5cm}{!}{\includegraphics{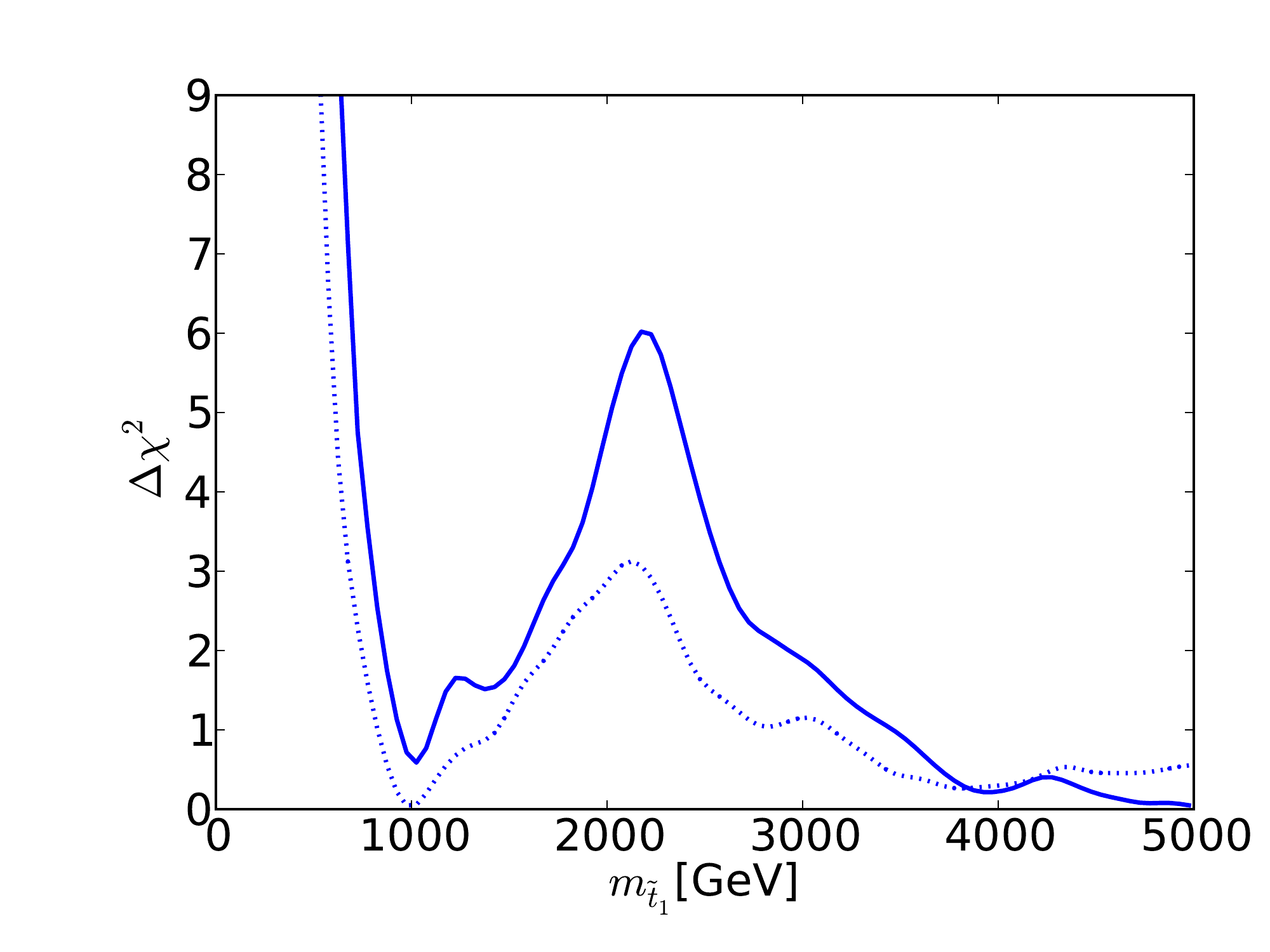}}
\resizebox{8.5cm}{!}{\includegraphics{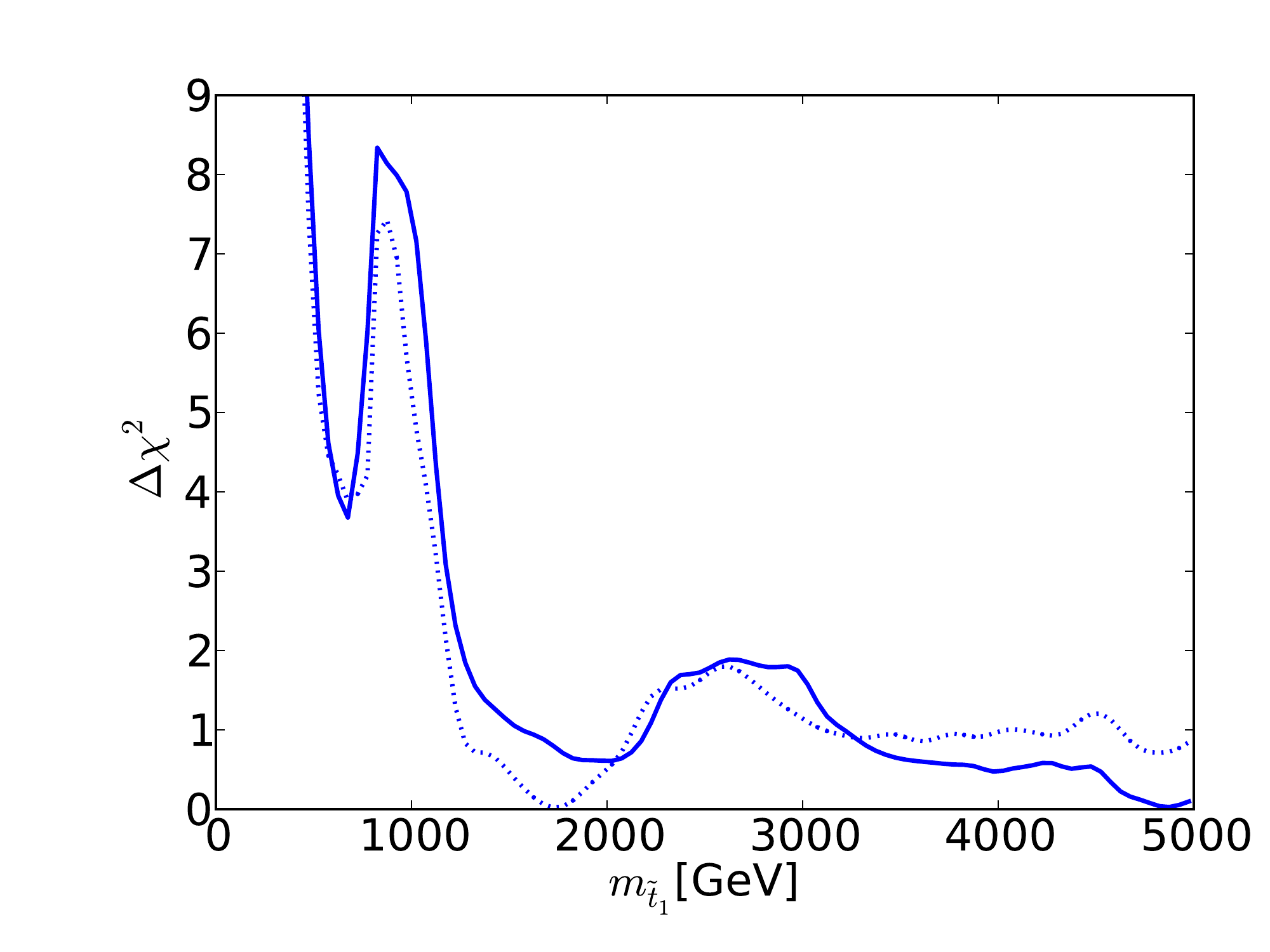}} \\
\resizebox{8.5cm}{!}{\includegraphics{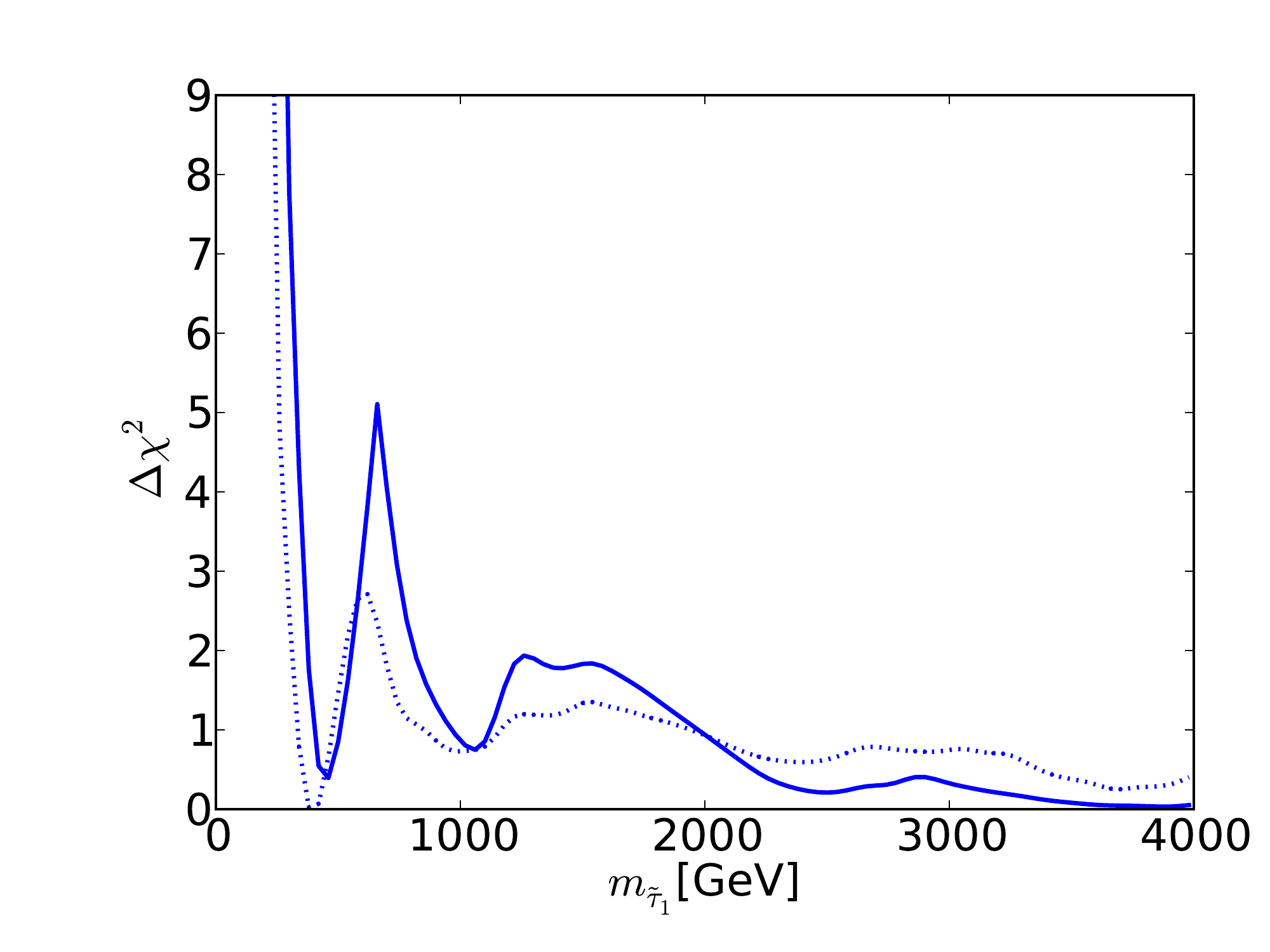}}
\resizebox{8.5cm}{!}{\includegraphics{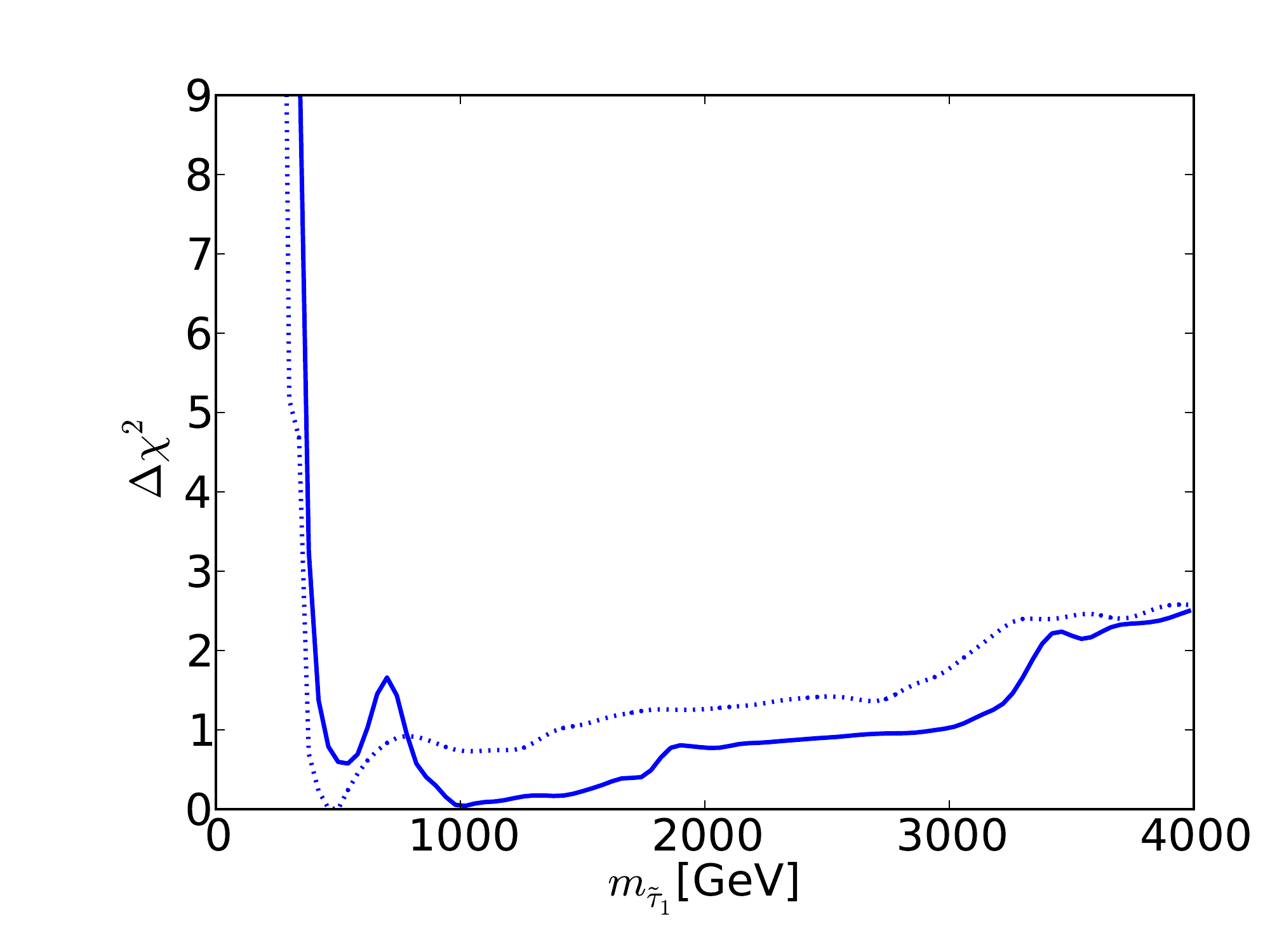}} 
%%%%%%%%%%%%%%%%%%%%%%%%%%%%%%%
%\vspace{-1cm}
\caption{The one-dimensional $\chi^2$ likelihood functions for $m_{\tilde t_1}$ (upper)
and $m_{\tilde \tau_1}$ (lower) in the CMSSM (left)
and the NUHM1 (right)~\protect\cite{mc9}. In each panel, the solid line is derived from a global analysis of the
present data, and the dotted line is derived from an analysis if the data set used
in~\protect\cite{mc8}, using the same implementations of the $m_h$ and dark matter scattering constraints.}
\label{fig:stop}
\end{figure*}

The $\chi^2$ function for the mass of the lighter stop squark $m_{\tilde t_1}$ in the CMSSM,
shown in the upper left panel of Fig~\ref{fig:stop}, exhibits
a local minimum at $m_{\tilde t_1} \sim 1000$~GeV
and a local maximum at $m_{\tilde t_1} \sim 2000$~GeV~\cite{mc9}.
On the other hand, the $\chi^2$ function for $m_{\tilde t_1}$ in the NUHM1, shown in the upper right panel of 
Fig~\ref{fig:stop}, exhibits a local maximum at $m_{\tilde t_1} \sim 1000$~GeV
and a local minimum at $m_{\tilde t_1} \sim 2000$~GeV, followed by another
local maximum at $m_{\tilde t_1} \sim 2600$~GeV.

%%%%%%%%%%%%%%%%%%%%%% F I G U R E %%%%%%%%%%%%%%%%%%%%%%%%%%%%%%%%%%%
%%%%%%%%%%%%%%%%%%%%%% F I G U R E %%%%%%%%%%%%%%%%%%%%%%%%%%%%%%%%%%%

The lower panels of Fig~\ref{fig:stop} show the $\chi^2$ functions for the lighter stau in the
CMSSM (left and the NUHM1 (right). In both cases, we see that low masses are strongly
disfavoured, and that the $\chi^2$ functions are almost flat above 1000~GeV, with
local maxima at $m_{\tilde \tau_1} \sim 700$~GeV.

\begin{figure}[htbp!]
%\begin{center}
\resizebox{8.5cm}{!}{\includegraphics{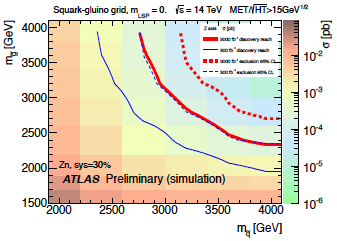}}
\caption{The physics reach of the LHC in the $(m_0, m_{1/2})$ plane provided by
searches for squarks and gluinos assuming that the LSP mass is negligible~\protect\cite{ATLAS-HL-LHC}. 
The different colours represent the production cross section at 14~TeV. The solid (dashed) lines
display the 5-$\sigma$ discovery reach (95\% CL exclusion limit) with 300/fb and 3000/fb respectively.
\label{fig:ATLASglsq}}
%\end{center}
\end{figure}

There is no indication of a preferred
supersymmetric mass scale, but one may set the following 95\% CL lower limits in GeV units~\cite{mc9}:
\begin{eqnarray}
m_{\tilde g} & > 1810~{\rm( CMSSM)}, & 1920~{\rm( NUHM1)} \, , \nonumber \\
m_{\tilde q_R} & > 1620~{\rm( CMSSM)}, & 1710~{\rm( NUHM1)} \, , \nonumber \\
m_{\tilde t_1} & > 750~{\rm( CMSSM)}, & 1120~{\rm( NUHM1)} \, \nonumber \\
m_{\tilde \tau_1} & > 340~{\rm( CMSSM)}, & 450~{\rm( NUHM1)} \, .
\label{masslimits}
\end{eqnarray}
For comparison, estimates of the supersymmetry discovery reach of the 
LHC with 14~TeV can be found in~\cite{ATLAS-HL-LHC},
e.g., the $(m_0, m_{1/2})$ plane displayed in Fig.~\ref{fig:ATLASglsq}. It ws estimated in~\cite{ATLAS-HL-LHC}
that the 5-$\sigma$ discovery reach for squarks and gluinos with 300/fb of high-energy luminosity
should be to $m_{\tilde g} \sim 3500$~GeV and $m_{\tilde q_R} \sim 2000$~GeV if $m_\chi \ll m_{\tilde g},
m_{\tilde q_R}$, and similar sensitivities are expected in the CMSSM and the NUHM1.
The discovery range with 3000/fb of luminosity
would extend a few hundred GeV further, so large parts of the CMSSM and NUHM1 parameter spaces will be
accessible in future runs of the LHC.

On the other hand, the lower panels in Fig~\ref{fig:stop} and the 95\% CL lower limits
on $m_{\tilde \tau_1}$ given in (\ref{masslimits}) suggest, within the CMSSM and NUHM1,
that the lighter stau and other sleptons may lie beyond the reach of a low-energy $e^+ e^-$
collider. However, it should be emphasized that this observation is necessarily model-dependent,
as there is no direct information on $m_{\tilde \tau_1}$. If the universality assumptions of the
CMSSM and the NUHM1 were to be modified appropriately, one might be able to explain
the $g_\mu - 2$ discrepancy as well as offering more hope for ${\tilde \tau_1}$ detection
in $e^+ e^-$ collisions.

\begin{figure*}[htb!]
%%%%%%%%%%%%%%%%%%%%%%%%%%%%%%%
\resizebox{8.5cm}{!}{\includegraphics{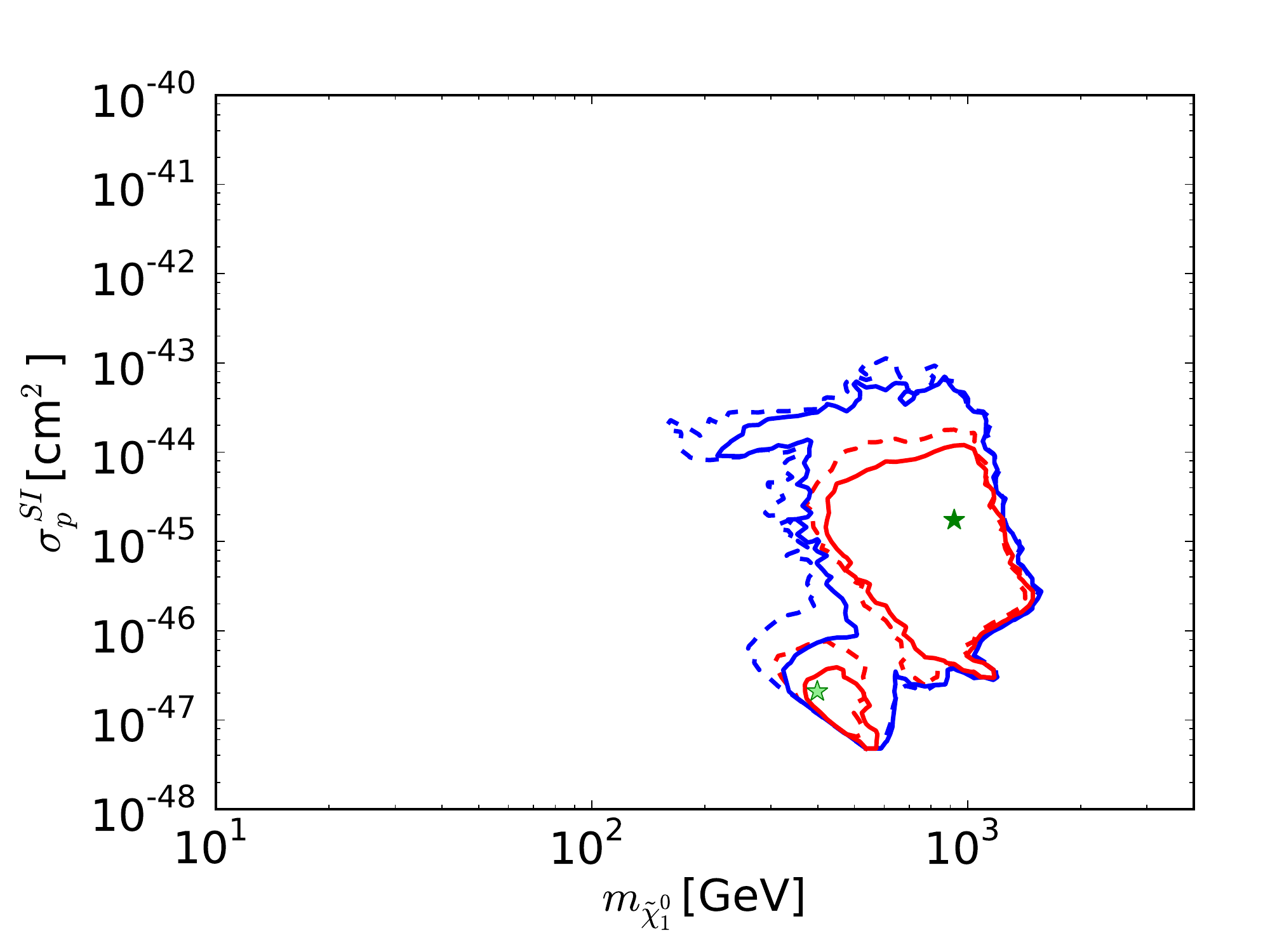}}
\resizebox{8.5cm}{!}{\includegraphics{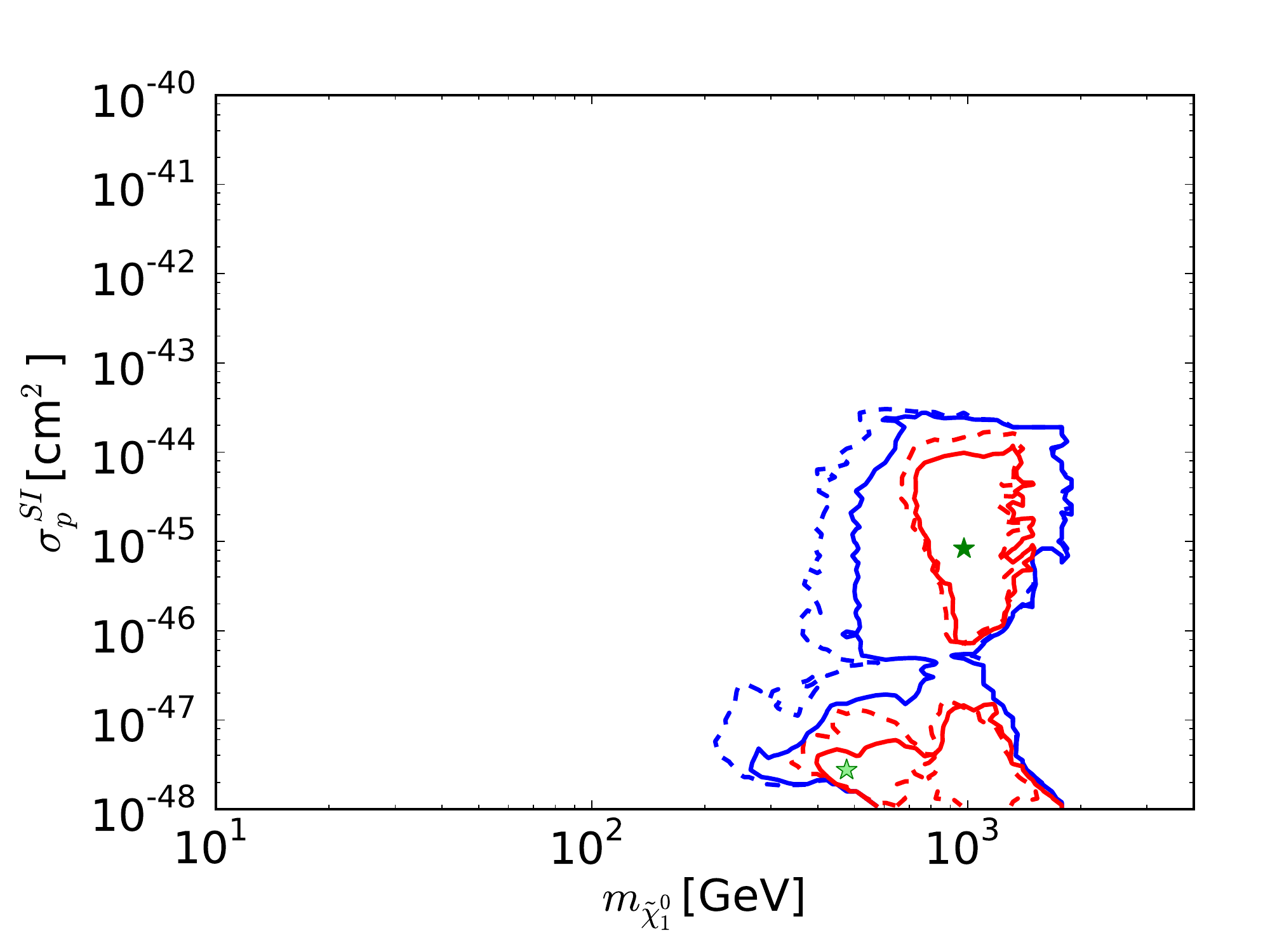}}
%%%%%%%%%%%%%%%%%%%%%%%%%%%%%%%
%\resizebox{8cm}{!}{\includegraphics{cmssm_mc9_mstop1_chi2.png}}
%\resizebox{8cm}{!}{\includegraphics{cmssm_mc9_mstau1_chi2.png}}
%%%%%%%%%%%%%%%%%%%%%%%%%%%%%%%
%\vspace{-1cm}
\caption{The $(m_\chi , \ssi)$ planes in the CMSSM (left)
and the NUHM1 (right)~\protect\cite{mc9}.
In both panels, the solid lines are derived from a global analysis of the
present data, and the dotted lines are derived from an analysis of the data used
in~\cite{mc8}, with the current implementations of the $m_h$ and \ssi\ constraints.
The red lines denote
the $\Delta \chi^2 = 2.30$ contours, the blue lines denote the $\Delta \chi^2 = 5.99$
contours in each case, and the filled (open) green stars denote the corresponding best-fit points.} 
\label{fig:CMSSMothers}
\end{figure*}

Fig.~\ref{fig:CMSSMothers} displays the $(m_\chi , \ssi)$ planes in the CMSSM (left)
and the NUHM1 (right), again with solid (dashed) lines representing the current analysis~\cite{mc9}
and the constraints of~\cite{mc8}, respectively, the red (blue) lines representing 68 (95)\%
CL contours, respectively, with the filled (open) green stars denoting the corresponding best-fit points.
We see that values of \ssi\ in range $10^{-47} \lsim \ssi\ \lsim 10^{-43}$~cm$^2$ are allowed 
in the CMSSM at the 95\% CL, though the best-fit point yields $\ssi\ \lsim 10^{-46}$~cm$^2$.
In the NUHM1, the range of \ssi\ preferred at the 68 and 95\% CL extends to lower values $\lsim 10^{-48}$~cm$^2$,
whilst the best-fit point yields $\ssi\ \sim 10^{-45}$~cm$^2$, higher than the CMSSM best-fit value.
These global fits indicate that \ssi\ may lie considerably below the current upper limit from
the LUX experiment~\cite{LUX}, though significantly above the level of the background from neutrino scattering,
and hence potentially accessible to future experiments searching for the scattering
of astrophysical dark matter.

%%%%%%%%%%%%%%%%%%%%%% F I G U R E %%%%%%%%%%%%%%%%%%%%%%%%%%%%%%%%%%%
%%%%%%%%%%%%%%%%%%%%%% F I G U R E %%%%%%%%%%%%%%%%%%%%%%%%%%%%%%%%%%%

There have been several claims to have observed signatures of the scattering
of relatively low-mass dark matter particles, which could not be accommodated
within the class of universal models discussed here. Moreover, these claims were not easy to reconcile
with other negative results, e.g., from XENON100, and seem now to have been ruled out by the first results
of the LUX experiment~\cite{LUX}. Likewise, there are various claims to have observed what might be indirect
signatures of annihilations of astrophysical dark matter particles that are also difficult to
accommodate within the class of models discussed here, and will not be discussed further.

\section{Alternative Approaches}

The above results were in the CMSSM and NUHM1 frameworks, and are quite specific
to those models. This Section contains some discussions of other models and
proposals for model-independent analyses of LHC data.

\subsection{mSUGRA}

\begin{figure}
\begin{center}
%\begin{tabular}{c c}
\resizebox{8.5cm}{!}{\includegraphics{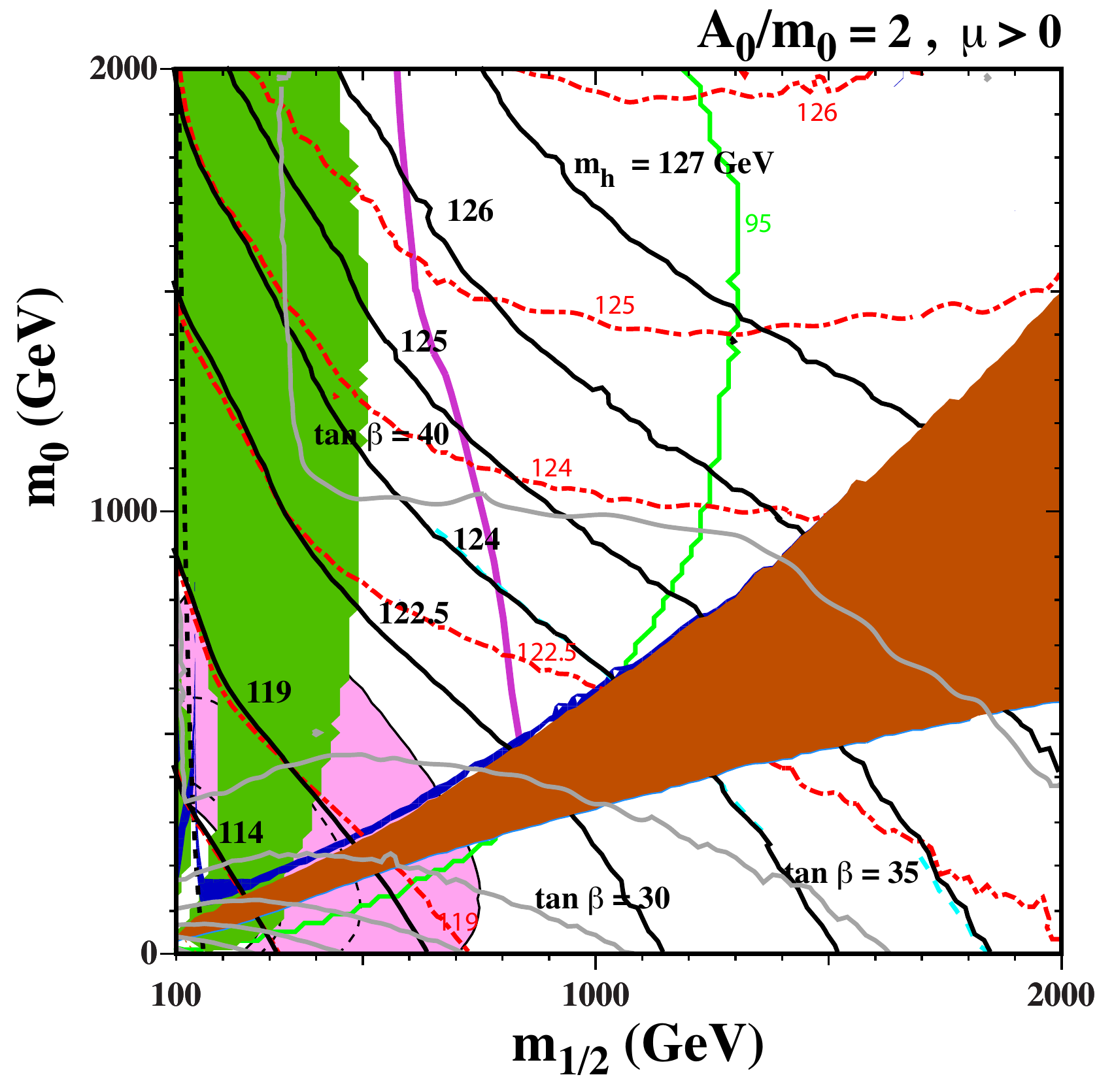}}
%\mbox{\epsfig{file=m0M_mA_30_2.5_1000v4.eps,height=7cm}} \\
%\end{tabular}
\end{center}   
%\begin{center}
%\begin{tabular}{c c}
%\mbox{\epsfig{file=mum0_Ms_10_2.5_1000v4.eps,height=7cm}} &
%\mbox{\epsfig{file=mumAs_10_2.5_1000_1000v4.eps,height=7cm}} \\
%\end{tabular}
%\end{center}   
\caption{\label{fig:mSUGRA}
The $(m_{1/2}, m_0)$ plane in a mSUGRA model with
$A_0/m_0 = 2$~\protect\cite{EHOW+}. In addition to the line and shade conventions used in
Fig.~\ref{fig:CMSSM30}, the values of $\tan \beta$ derived from the
electroweak vacuum conditions are shown as solid grey contours.
}
\end{figure}

As already mentioned, mSUGRA is a more restrictive framework than the CMSSM,
since the gravitino mass is equal to the scalar mass: $m_{3/2} = m_0$, and the trilinear and bilinear
soft supersymmetry-breaking parameters are related: $A_0 = B_0 + m_0$. The former relation
restricts the part of the $(m_{1/2}, m_0)$ plane in which the lightest neutralino is the LSP, 
and the second relation allows the value of $\tan \beta$ to be fixed at each point in the
$(m_{1/2}, m_0)$ plane by the electroweak vacuum conditions. Fig.~\ref{fig:mSUGRA}
displays a typical mSUGRA $(m_{1/2}, m_0)$ plane for the particular choice $A_0/m_0 = 2$~\cite{EHOW+}.
The same conventions as in Fig.~\ref{fig:CMSSM30} are used to represent the
experimental and cosmological density constraints, and the grey lines are contours of
$\tan \beta$. There is a (brown) wedge of the plane
where the LSP is the lighter stau, flanked by a neutralino LSP region at larger $m_0 = m_{3/2}$
and a gravitino LSP region at smaller $m_0 = m_{3/2}$. The ATLAS $\ETslash$ search is
directly applicable only in the neutralino LSP region, and would require reconsideration in the
gravitino LSP region. In addition, in this region there are important astrophysical
and cosmological limits on long-lived charged particles (in this case staus).
The (purple) ATLAS $\ETslash$ constraint intersects the (dark blue) dark matter coannihilation strip just above
this wedge where $m_{1/2} \sim 850$~GeV, and the (green) BR($B_s \to \mu^+ \mu^-$) constraint
intersects the coannihilation strip at $m_{1/2} \sim 1050$~GeV. The portion of the coannihilation
strip between this value and its tip at $m_{1/2} \sim 1250$~GeV is consistent with all the
constraints. In particular, in this section of the coannihilation strip the nominal
value of $m_h$ provided by {\tt FeynHiggs~2.10.0} is $\in (124, 125)$~GeV, compatible
with the experimental measurement within the theoretical uncertainties due to the 1-2 GeV shift
in $m_h$ found in the new version of {\tt FeynHiggs}, whereas the previous version would
have given $m_h < 124$~GeV.

\subsection{`Natural' Models}

\begin{figure}
\begin{center}
%\begin{tabular}{c c}
\resizebox{8.5cm}{!}{\includegraphics{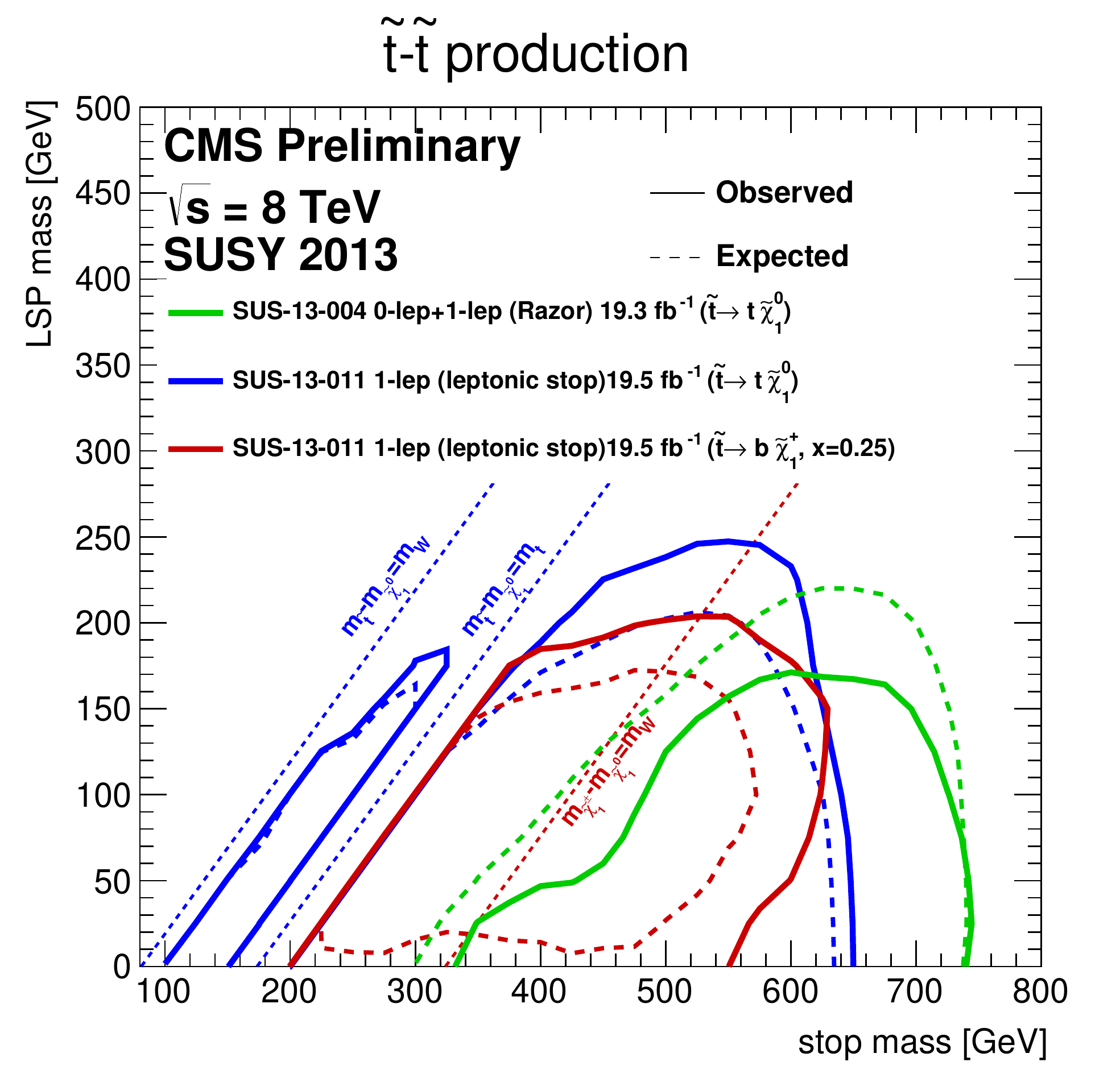}}
%\mbox{\epsfig{file=m0M_mA_30_2.5_1000v4.eps,height=7cm}} \\
%\end{tabular}
\end{center}   
%\begin{center}
%\begin{tabular}{c c}
%\mbox{\epsfig{file=mum0_Ms_10_2.5_1000v4.eps,height=7cm}} &
%\mbox{\epsfig{file=mumAs_10_2.5_1000_1000v4.eps,height=7cm}} \\
%\end{tabular}
%\end{center}   
\caption{\label{fig:stops}
Exclusion limits from stop searches by the CMS Collaboration~\protect\cite{CMSstop}.
}
\end{figure}

In view of the absence of supersymmetry in conventional jets + $\ETslash$
searches, the fact that the lighter stop squark ${\tilde t_1}$ is lighter than first- and 
second-generation squarks in many models (as we saw earlier in the cases
of the CMSSM and the NUHM1), and the fact that the naturalness (or fine-tuning)
argument applies most strongly to the stop, there have been many studies of
so-called `natural' models in which it is assumed that $m_{\tilde t_1} \ll m_{\tilde q_R}, m_{\tilde g}$.
Fig.~\ref{fig:stops} summarizes the results of dedicated stop searches by the CMS Collaboration~\cite{CMSstop}.
We see explicitly that the sensitivity of search depends on the stop decay mode
assumed as well as the LSP mass assumed, and should recall that in a realistic model 
stop decays may not be dominated by a single mode. So far, the dedicated stop
searches do not impinge significantly on the parameter spaces of the CMSSM and
the NUHM1, but this may change in the future.

\subsection{Simplified Models}

\begin{figure}
\begin{center}
%\begin{tabular}{c c}
\resizebox{8.5cm}{!}{\includegraphics{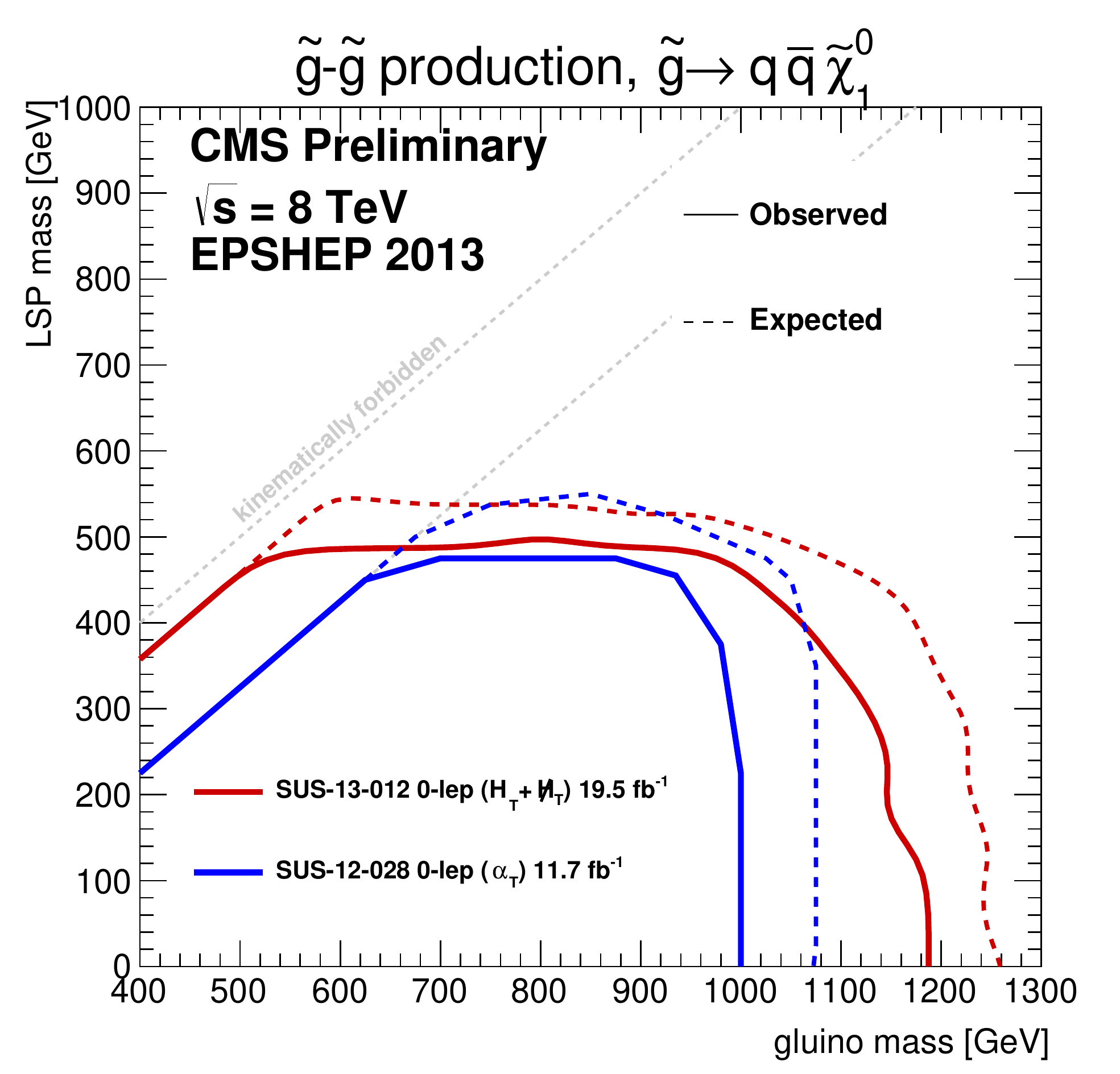}}
%\mbox{\epsfig{file=m0M_mA_30_2.5_1000v4.eps,height=7cm}} \\
%\end{tabular}
\end{center}   
%\begin{center}
%\begin{tabular}{c c}
%\mbox{\epsfig{file=mum0_Ms_10_2.5_1000v4.eps,height=7cm}} &
%\mbox{\epsfig{file=mumAs_10_2.5_1000_1000v4.eps,height=7cm}} \\
%\end{tabular}
%\end{center}   
\caption{\label{fig:SM}
Exclusion limits from searches by the CMS Collaboration in the
simplified model topology ${\tilde g} {\tilde g} \to {\bar q} q {\bar q} q \chi \chi$~\protect\cite{CMSstop}.
}
\end{figure}

Another approach has been to benchmark supersymmetric searches by
assuming simplified models in which some specific cascade signature
is assumed to dominate sparticle production and decay at the LHC. For
example, it might be assumed that the gluinos are much lighter than all 
the squarks and decay dominantly into ${\bar q} q \chi$ final states.
Fig.~\ref{fig:SM} shows the exclusion limits obtained by the CMS
Collaboration from a search for pair-production of gluinos in this
heavy-squark limit followed
by decays into ${\bar q} q \chi$ final states with 100\% branching ratios~\cite{CMSstop}.
We see that this search also does not reach the 95\% CL lower
limits in the CMSSM and the NUHM1 that were discussed earlier.
We also note that such simplified models are in general over-simplified,
in that typical branching ratios are $< 100$\%, on the one hand, and
realistic models may be tackled simultaneously using several signatures
in parallel. A possible way forward building on the simplified model
approach may be to parameterize a realistic model in terms of the
probabilities with which specific model signatures occur and combine
different signatures with a `mix and match' approach to obtain the overall
sensitivity to that model~\cite{Sakurai}.

\subsection{Combining Searches}

\begin{figure}
\begin{center}
%\begin{tabular}{c c}
\resizebox{8.5cm}{!}{\includegraphics{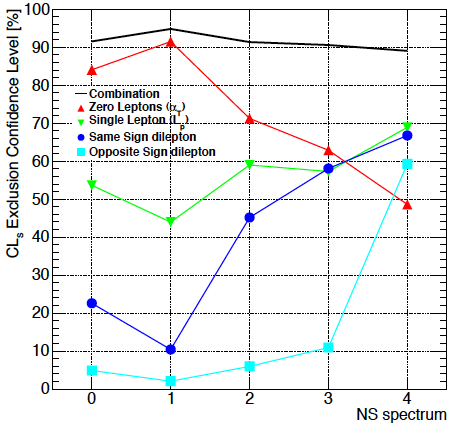}}
%\mbox{\epsfig{file=m0M_mA_30_2.5_1000v4.eps,height=7cm}} \\
%\end{tabular}
\end{center}   
%\begin{center}
%\begin{tabular}{c c}
%\mbox{\epsfig{file=mum0_Ms_10_2.5_1000v4.eps,height=7cm}} &
%\mbox{\epsfig{file=mumAs_10_2.5_1000_1000v4.eps,height=7cm}} \\
%\end{tabular}
%\end{center}   
\caption{\label{fig:BM}
The confidence levels for excluding a class of `natural-like'
supersymmetric models by combining searches at
7~TeV for several different topologies:
$\ETslash$ + jets without leptons, with a
single lepton and with same- and opposite-sign dileptons~\protect\cite{BM}.}
\end{figure}

An interesting step in this direction was taken in~\cite{BM}, where
it was shown that certain combinations of searches yield a sensitivity
to a class of models that is almost independent of the specific
parameters of the model within that class. The idea here was to
combine searches for $\ETslash$ + jets without leptons, with a
single lepton and with same- and opposite-sign dileptons, and apply
them to a class of `natural-like' supersymmetric spectra. As can be
seen in Fig.~\ref{fig:BM} where this approach was applied to 7~TeV
data, the confidence level with which a particular set of gluino,
third-generation squark and LSP masses
($m_{\tilde g} = 1$~TeV, $m_{\tilde q_3} = 700$~GeV, $m_\chi = 100$~GeV)
could be excluded was found to be essentially independent of
other details of the spectrum and associated branching ratios.

\subsection{Monojet Searches}

In all the above searches, the production and cascade decays
of heavier supersymmetric particles were considered. A different
approach, which aims to be more model-independent, is to
look directly for pair-production of LSPs $\chi$ with the
signature of an accompanying monojet (due predominantly
to initial-state gluon radiation) or electroweak boson ($\gamma$,
$W^\pm$ or $Z^0$). The idea was to use such searches to
constrain higher-dimensional operators that could also mediate
the scattering of astrophysical dark matter. In particular, it was 
hoped that this approach would clarify the confusion that existed
for a long time about possible experimental hints for low-mass
cold dark matter particles.

\begin{figure*}[htb!]
%%%%%%%%%%%%%%%%%%%%%%%%%%%%%%%
\resizebox{17cm}{!}{\includegraphics{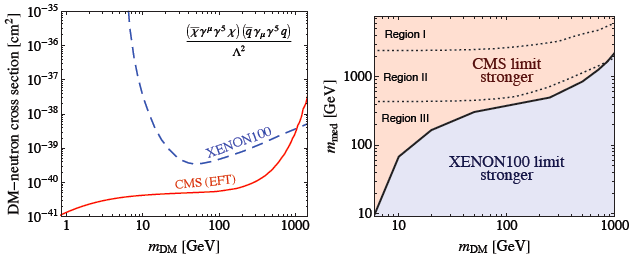}}
%%%%%%%%%%%%%%%%%%%%%%%%%%%%%%%
%\resizebox{8cm}{!}{\includegraphics{cmssm_mc9_mstop1_chi2.png}}
%\resizebox{8cm}{!}{\includegraphics{cmssm_mc9_mstau1_chi2.png}}
%%%%%%%%%%%%%%%%%%%%%%%%%%%%%%%
%\vspace{-1cm}
\caption{It is shown in the left panel that, in the effective field theory (EFT) approximation,
monojet searches are more sensitive than the XENON100 search for a spin-dependent
dimension-6 interaction of the form 
$({\bar \chi} \gamma_\mu \gamma_5 \chi)({\bar q} \gamma_\mu \gamma_5 q)/\Lambda^2$.
However, the right panel shows that this conclusion depends on the mass controlling the form factor
of the dimension-6 interaction~\protect\cite{BDM}.} 
\label{fig:BDM}
\end{figure*}

This approach looks promising for the case of spin-dependent
dark matter scattering via an effective dimension-6 operator of the form
$({\bar \chi} \gamma_\mu \gamma_5 \chi)({\bar q} \gamma_\mu \gamma_5 q)/\Lambda^2$,
as seen in the left panel of Fig.~\ref{fig:BDM}~\cite{BDM}. However, one should
remember that the kinematics of dark matter scattering (which has a very
small space-like momentum transfer) and pair-production (where the momentum
transfer is time-like and $> 4 m_\chi^2$). This raises the possibility that there
may be a non-trivial form factor for the effective operator, which could suppress
the sensitivity in the LHC searches for monojets, etc.. The right panel of
Fig.~\ref{fig:BDM} illustrates the potential importance of this effect. Whereas
the LHC limit appears stronger than the XENON100 limit in the effective
field theory (EFT) limit (left panel), we see that the XENON100 limit may
actually be stronger, depending on the details of the theory underlying the
EFT model~\cite{BDM}. That said, this approach is an interesting supplement to
more conventional $\ETslash$ + jets searches, and may play an
increasingly important r\^ole in searches for supersymmetry and other new physics
when the LHC restarts at high energy.

%%%%%%%%%%%%%%%%%%%%%% F I G U R E %%%%%%%%%%%%%%%%%%%%%%%%%%%%%%%%%%%
%%%%%%%%%%%%%%%%%%%%%% F I G U R E %%%%%%%%%%%%%%%%%%%%%%%%%%%%%%%%%%%

\section{Summary and Prospects}

The first run of the LHC leaves a bittersweet taste in the mouths of
high-energy physicists. On the one hand, the ATLAS and CMS Collaborations
have discovered a Higgs boson, an experimental Holy Grail since it was first
postulated in 1964. On the other hand, they have found no trace of any other new
physics, in particular no sign of supersymmetry. However, the appearance of
an apparently elementary Higgs boson poses severe problems of naturalness
and fine-tuning, so theorists should rejoice that they have new challenges to
meet. Supersymmetry still seems to the present author to be the most promising
framework for responding to these challenges, and I argue that the LHC
measurements of the low mass and Standard Model-like couplings of the Higgs boson
provide additional circumstantial arguments for supersymmetry.

The LHC searches for supersymmetry, the Higgs mass, the measurement of
\bsmm\ and other experiments, notably those on dark matter, can be combined
in global fits to the parameters of specific supersymmetric models~\cite{mc9,others}. The two
examples discussed here are the CMSSM and the NUHM1: analyzing models
with more parameters in an equally thorough way would be far more
computationally intensive. Results of global fits to the CMSSM and the NUHM1,
including best-fit points, regions preferred at the 68\% CL and allowed at the
95\% CL have been presented in this paper, as well as 95\% CL lower limits
on some sparticle masses. Within these models, there are reasonable prospects for discovering 
supersymmetry at the LHC at higher energy, as well as for observing the
scattering of astrophysical dark matter.

Various alternative approaches to supersymmetry phenomenology have also
been discussed, including `natural' models, simplified models, combined analyses
of benchmark signatures, and searches for monoboson events. Although none of
these impinges significantly on the CMSSM and NUHM1 parameter spaces, all of them are
likely to play greater r\^oles in future studies of supersymmetry at the LHC at higher
energies, particularly as interest broadens to a wider range of models.

We await with impatience the advent of high-energy LHC running with increasing luminosity.

\section*{Acknowledgements}

I thank fellow members of the MasterCode Collaboration, particularly
Oliver Buchmueller, Sven Heinemeyer, Jad Marrouche, Keith Olive and Kees de Vries
for many discussions on this subject.
This work was supported in part by the London Centre for Terauniverse Studies (LCTS),
using funding from the European Research Council 
via the Advanced Investigator Grant 267352.

\end{document}